\documentclass[traditabstract,letter]{aa} 
\usepackage{longtable}
\usepackage{amsfonts}
\usepackage{rotating}
\RequirePackage{amssymb}
\RequirePackage{latexsym}

\usepackage{natbib}

\usepackage[german,english]{babel}
\usepackage{graphicx}
\usepackage{epsfig}
\usepackage{xspace}
\usepackage{epsf}
\usepackage{epstopdf}

\usepackage{color}
\usepackage{epstopdf}
\usepackage[bottom]{footmisc}

\def\simpropto{\lower.2ex\hbox{$\; \buildrel \propto \over \sim \;$}}
\def\ltsim{\lower.5ex\hbox{$\; \buildrel < \over \sim \;$}}
\def\gsim{\lower.5ex\hbox{$\; \buildrel > \over \sim \;$}}
\def\ltsim{\lower.5ex\hbox{$\; \buildrel < \over \sim \;$}}
\def\gtsim{\lower.5ex\hbox{$\; \buildrel > \over \sim \;$}}

\def\ltsim{ \,{}^<_\sim\, } 

\newcommand{\vg}{\color{cyan}}

\usepackage{txfonts}
\begin{document}
 \title{Jet interactions with a giant molecular cloud in the Galactic centre  and ejection of hypervelocity stars}

\author{ Joseph Silk\inst{1,2,3}
\and Vincenzo Antonuccio-Delogu\inst{4}
\and Yohan Dubois\inst{1}
\and Volker Gaibler\inst{5}
\and Marcel R. Haas \inst{6}
\and Sadegh Khochfar\inst{7}
\and Martin Krause\inst{8,7,9}
}
\institute{Institut d'Astrophysique de Paris, CNRS, 98bis bd Arago, 75014 Paris, France 
\and Department of Physics and Astronomy, The Johns Hopkins University
Homewood Campus, Baltimore MD 21218, USA 
\and BIPAC, Department of Physics, University of Oxford, Oxford OX1 3RH, UK  \\
\email{silk@astro.ox.ac.uk}
\and INAF - Osservatorio Astrofisico di Catania, Via S. Sofia 78, Catania, 
\& Scuola Superiore di Catania, Via Valdisavoia 9, Catania, I-95123, Italy 
\and {Universit\"at Heidelberg, Zentrum f\"ur Astronomie, Institut f\"ur Theoretische Astrophysik,
      Albert-Ueberle-Str. 2, 69120 Heidelberg, Germany } 
\and STScI,  3700 San Martin Drive, Baltimore, MD, 21218, USA 
\and Max-Planck Institut f\"ur extraterrestrische Physik, PO Box 1312, D-85478 Garching, Germany 
\and Excellence Cluster Universe, Technische Universit\"at M\"unchen,
     Boltzmannstrasse 2, 85748 Garching
 \and
         Geneva Observatory, University of Geneva, 51 Chemin des Maillettes, 1290 Versoix, Switzerland
}


   \date{}

 
  \abstract
   {
The hypervelocity OB  stars in the Milky Way Galaxy were ejected from the central regions some 10-100 million years ago. We argue that these stars, {as well as many more abundant bound OB stars in the innermost few parsecs,}  were generated by the interactions of an AGN jet from the central black hole with a dense molecular cloud. Considerations of the associated energy and momentum injection 
 have broader implications for the possible origin of the Fermi bubbles and for the enrichment of the intergalactic medium.
}


   \maketitle
%

\section{Introduction}
Hypervelocity stars (HVS) in our galaxy defy explanation.  With velocities directed outward from the centre of the galaxy of  between 300 and 1000 km/s, these young OB stars cannot be explained by ejection from binaries or via binary encounters with the central black hole.  

Most hypervelocity stars seem to be  the relics of  something more
exotic than binary ejections. 
The more massive the star, the higher the run-away fraction.
Since a significant fraction are O stars of$ \gsim  40 \rm M_\odot ,$ this means the event that generated them was relatively recent.  Binary star scattering by a central star cluster fails
by at least two orders of magnitude
to account for their frequency \citep{2012arXiv1202.2356P}.
Binary ejection in supernovae seems unlikely given that 
their orbits are consistent with coming from the nucleus of the Milky Way Galaxy (MWG) and the travel times 
of hypervelocity  B supergiants, around $3-4 \rm M_\odot$, at 50-100 kpc from the Galactic centre (GC) and with  
velocities$\gsim  300 $km/s,  are constrained   to 60-200 Myr
\citep{2012ApJ...754L...2B}.

More exotic possibilities have been investigated. The
  Milky Way's central supermassive black hole might itself have a
  companion black hole, which could kick off the stars at high speed 
\citep{2003ApJ...599.1129Y,2006MNRAS.372..174B,2008ApJ...686..432S}.
The main difficulties with this scenario are that 
 the velocities could  come
out to be too high \citep{2008ApJ...686..432S}, and that  if dynamical friction
with the stars in the dense Galactic centre is taken into account, a
continuous supply of intermediate mass black holes inspiralling into
the GC must be invoked, because each one may only eject stars for a
timescale of $10^6$~yrs \citep{2006MNRAS.372..174B}.
About 100 HVS (B stars of $3-10\rm M_\odot$) are generated in the halo over a typical propagation time of 100 Myr, or at a mean ejection rate of
 ~1/Myr ejected.
We infer that the ejection most likely was spread over $10^6$ to
$10^8$ yrs because of the spread in distances travelled  and lifetime
considerations.

Recurrent AGN activity is an established piece of galaxy
  evolution, in contrast to recurrent accretion of intermediate mass
  black holes. Hence
 if there is no other convincing 
explanation, it is useful to estimate whether  the mechanism of origin of hypervelocity stars
may  be due to positive feedback on molecular gas induced by  jet
interactions generated by past explosions from our central  {{supermassive black hole (SMBH)}.

AGN jets have  recently been shown via 3-D simulations to  overpressure clouds and induce star formation in  gas-rich central disks
\citep{2011arXiv1111.4478G}.  Another recent study of AGN jets interacting with an cloudy interstellar medium   demonstrates that $10\%$ or more of the jet energy effectively accelerates the gas clouds to escape velocity over a few  tens of millions of years
\citep{2012arXiv1205.0542W}.
 Neither of these studies includes cloud self-gravity, and one must await new studies before reaching any definitive conclusions, for example on cloud survival. However it is nevertheless  useful to point out that there may be a local counterpart of jet-induced gas flows and induced star formation, for which the energetics and efficiencies seem to work out surprisingly well.

Our own central black hole currently has a very low accretion rate, but this may not have been the case in the past. Indeed, { jet activity has recently been claimed from radio data \citep{2012arXiv1208.1193Y}. This leads  the possibility that an eus (AGN)-like phenomenon may have repeatedly occurred in our own Galactic centre. 
Indeed, the recently discovered Fermi bubbles in the central regions of the Galaxy may provide evidence for an  outburst some 10 million years ago, provided reacceleration of energetic electrons occurred in the associated shocks \citep{2010ApJ...724.1044S, 2011arXiv1103.0055G, 2011MNRAS.415L..21Z}.  
Hypervelocity  B stars suggest an event some 100 million years ago. 


Our starting point is  that of momentum considerations. There is an intriguing  coincidence between the momentum flow from the GC and that in hypervelocity stars that connects, we believe, to an outstanding problem in galaxy formation theory. {The point here is that production of HVS is inefficient. Most of the stars formed by jet-induced  overpressuring of GC gas clumps are inevitably at relatively low velocity. These are tracers however of the same mechanism that  generated the rarer HVS.} Hence we first describe star formation in the inner 
few parsecs.

\section {Star formation in the Galactic centre}

The Milky Way Nuclear Star Cluster has a radius of 5 pc and a mass of $2-3. 10^7 \rm M_\odot$  \citep{2002A&A...384..112L}.
It  has several components, including 
the Sgr A* star cluster which contains  $\sim  200$ young stars within the central parsec.  There is also disk structure, with a main and an inclined disk at 0.8-12 arcsec (1 arc-sec = 0.04pc)  of massive stars formed 6 Myr ago, as well as a more isotropic distribution of older  ($> 1$ Gyr old) stars containing $90\%$  of the stellar mass within  the central  parsec \citep{2011ApJ...741..108P}.

The IMF of the WR, O, B stars in the central disks (in the innermost 4000 AU, 0.8-12 arc-sec)  is top-heavy \citep{2010ApJ...708..834B}.
 The WR and O stars  formed coevally about 6 Myr ago. In addition there are late B (less massive) stars more isotropically distributed. Closer in there is an  old star cluster  with a standard IMF centred on Sgr A*. The B stars beyond  12 arc-sec also have a standard IMF.  
 There could be a correction for less massive stars, possibly  as large as 10, in the  central arc-sec. 
 If the central OB stars (within 1 arc sec) were also  formed by the jet interaction ~ 10 Myr ago, the HVS ejection efficiency is inferred to be of order  1-10$\%$. 

Further out, the Arches cluster alone has a present day kinematic mass  $1.5 \times10^4 \rm M_\odot$ \citep{2011arXiv1112.5458C}.
It is 26 pc from the GC and only 2 Myrs old. 
The precursor molecular cloud mass of an Arches-like young massive star cluster    is observed to be at a  typical scale of 3 pc and mass of $10^5 \rm M_\odot$, see   \cite{2012ApJ...746..117L}. In our model, the jet-driven bubble radius accelerates and compresses gas out to $\sim 30$  pc, where it can continue to induce star formation.

At  $10\%$ star formation efficiency (SFE), we  would expect $10^5\rm M_\odot$ in stars. In fact we see some  $3 .10^4 \rm M_\odot$  in low mass stars within 0.1 pc by using the integrated light  at HST resolution of the central arc-second of the GC  \citep{2012ApJ...744...24Y}.
\subsection{{A giant molecular cloud around the central SMBH}}
At high redshift, the specific star formation rate is observed to be high. This is due either to an increase in the efficiency of star formation related perhaps to the increase in gas supply or to a new mode of star formation. The former interpretation, if coupled to the  metallicity dependence of molecular gas formation, fails to simultaneously fit low and high redshift specific star formation rates 
\citep{2011MNRAS.410L..42K,  2012ApJ...753...16K}.

The occurrence of a new mode is plausibly related to the rapid rise in AGN activity, that parallels star formation rate histories, by $z\sim 2.$ AGN interactions with gas-rich disks can provide positive as well as negative feedback, with  positive feedback naturally augmenting the specific star formation  rate, as in  \cite{2011arXiv1111.4478G}. 
The latter study shows how the pressure enhancements associated with a powerful jet can pressurise the gas-rich disk and thereby trigger and accelerate the star formation rate. 
One consequence is ejection of gas clumps. We argue here that simple considerations of the momentum transfer and energetics support the case that stars forming in the ejected gas clumps can account for the hypervelocity stars, if the last such jet episode occurred some ten million years ago at the GC, and followed other similar episodes with an appropriate duty cycle determined by replenishment of the gas reservoir.  



Dense molecular gas is mapped at the GC.   Some is infalling and some is undergoing tidal disruption, but a significant fraction ends up  in a central circumnuclear gas disk  at 1.5-4 pc from the GC \citep{ 2012arXiv1207.6309L}.  
 The mass of the inner giant molecular cloud  (GMC) amounts to $10^{6-7} \rm M_\odot$. 
 The residual gas  mass depends on  the star formation efficiency and on the assumed IMF.  
 Gas replenishment will guarantee repeated episodes of AGN activity and star formation.
If   we assume that non-axisymmetric gravitational instabilities drive GMC formation and infall, the gas replenishment time is of order $3. 10^7\rm yr.$ 
The inferred duty cycle for AGN activity and nuclear star formation is a few percent. For simplicity, we assume the cloud
to be spherical, 
although the physical mechanisms at work are expected to similarly apply for a wide range of geometries.

{\subsection{The role of AGN in cloud disruption}} 
Let us consider momentum flux balance for a cloud of mass $M_{cloud}$ and internal velocity dispersion $\sigma$: $ d(M_{cloud}\sigma)/dt=\dot M_w v_w$ or 
$M_{cloud}=f_Ef_gL_E{ t_{dyn}({ c \sigma})^{-1}},$
where  a geometrical factor $f_g \sim 1$ allows for the jet inefficiency in driving a quasispherical bow shock, $L_E$ is the Eddington luminosity, $f_E$ is the Eddington ratio ($L_{AGN}/L_E$) and $v_w\sim (0.1-0.3)c$ is the initial wind velocity for a wind of mass outflow rate $M_w.$
During the active phase of the AGN, we take  $f_E=0.1f_{E,0.1}$ with 
$f_{E,0.1}\sim 1,$  and one can disrupt some $10^6\rm M_\odot.$ The bow shock radius $r_s$ is set by $4\pi \rho_g r_s^2 v_{inf}^2 =\dot M_w v_w,$
or  
$r\approx 2 \rm pc \left[ f_{E,0.1}f_g M_{BH,4}n_{5}^{-1}v_{inf,50}^{-2}\right]^{1/2},$
where 
$ v_{inf}=50{\rm km\,  s^{-
1}}v_{inf,50}, $  $n=10^5n_5 \rm cm^{-3}, $ and 
$M_{BH}=4{ \times}10^6M_{BH,4}\rm M_\odot $

This is a conservative estimate, since the increased area of the thermal blast wave means that the jet-induced ram pressure  plus ambient medium shocked gas thermal  pressure exceeds  the jet input momentum by one or even two orders of magnitude: {\it eg.}  $f_g \sim 100$ at $\sim 10^6$ yr, depending on jet evolution \citep{2012arXiv1205.0542W}. 
The swept-out radius is of order
a few parsecs, comparable to the molecular cloud scale. We assume the cloud feeds the AGN (and forms stars) over an initial dynamical time, and  so the cloud should  be disrupted by the AGN over a time-scale 
$\propto r^4$, 
of order $10^{6}$ yr. A more detailed jet-driven bubble model is described below.

\subsection{Star formation triggering {by  AGN}}

 BH feeding and triggered star formation occur simultaneously. There is overwhelming evidence that AGN are capable of quenching star formation. This is certainly likely to be the case for the GMC within  a few parsecs of the central black hole. 
 However there is also expected to be star formation within the self-gravitating accretion disk that directly feeds the SMBH via both fragmentation and triggering.
%
We expect the triggered  nuclear  star formation rate to be regulated by the enhanced pressure and to be 
proportional to the square root of the pressure
\citep{2009ApJ...700..262S}.  
%

  For a SFE of 0.02 per dynamical time in  a $10^6\rm M_\odot$ GMC, one forms $ 2.10^4\rm M_\odot $ of stars per dynamical time,
$10^5 \rm yr, $ 
or 20 times the nuclear star formation rate (SFR). 
The mass  fraction of {this} GMC  that forms stars over a cloud  lifetime  is of order $\sim 1\%.$  The high star formation rates observed near AGN
motivate us to assert that the star mass fraction formed is elevated,
to say $\sim 10\%.$ 
The enhanced SFR is justified if the
pressure of the central cloud is elevated by a factor $\sim 100$
compared to nearby GMCs, as is the case if the cloud is
self-gravitating and of size 10 pc and  mean density $\sim 10^{4-5} 
\rm cm^{-3}.$
%

A top-heavy IMF is  motivated observationally 
for the star-forming disks at the GC \citep{2010ApJ...708..834B}.  If the cloud continues to  form stars for $\sim 10$ cloud free-fall times, it  forms some 20\% of its mass in stars over its lifetime, or $2.10^5\rm M_\odot$ in stars.  There are of order $10^4$ OB stars for a top-heavy IMF.

%
%




%
   \begin{figure}
   \centering
   \includegraphics[width=0.35\textwidth,angle=-0]{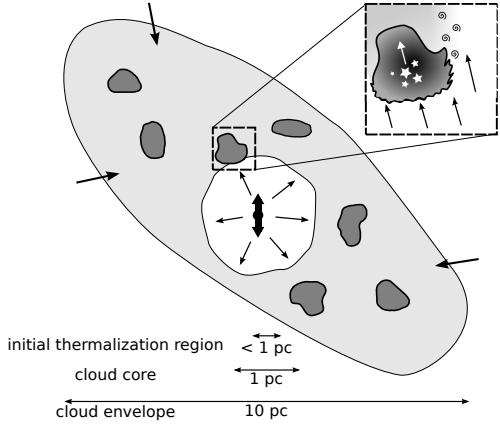}
      \caption{ Schematic model of jet/blast wave - cloud 
interaction in central 10 pc. The jet inflates a cocoon which punches into the GMC.
The cloud is subjected to 
the wind arising from the systematic expansion of the cocoon, which has the 
dual effect of compressing its outer layers,  thereby triggering star formation,  and accelerating some  the newly formed  stars 
to become hypervelocity stars.}
   
         \label{jet}
   \end{figure}
%
    
\section {The model} 

%
%
%
%
%
%
%
{The simulations 
  of the interaction of a jet with a
  gas-rich disk galaxy \cite{2011arXiv1111.4478G}  show that radio lobes  develop and drive a
 quasispherical blast wave, as long as the jet length is smaller
than the vertical scale height of the disk. The jet then escapes vertically and
the sideways expansion of the blast wave stalls. HVS form in the shell
at the sites of pre-existing over-densities, as long as the shell expands
rapidly. After breakout, more stars form with smaller velocities.

This picture may also apply to the disky cloud around our 
galaxy's SMBH.  In this case, the jet should have a total power $L$
of a fraction of the Eddington luminosity of $L_\mathrm{E}=5.10^{44}$~ergs/s 
(for mass $4.10^6M_\odot$). The jets will collimate and develop
lobes from around the inner scale $L_1$ \citep[e.g.][]{2012arXiv1206.1778K}, given by
$L_1 \approx 0.01$~pc~$ (L/L_\mathrm{E})^{1/2}
(\rho_0 /10^5\,m{\mathrm p})^{-1/2} (v_\mathrm{j}/c)^{-3/2}$, where the 
central cloud density is $\rho_0$, and the jet velocity is $v_\mathrm{j}$.

Within a molecular cloud centred on the SMBH, a jet
should develop lobes and bow shocks just as in the more familiar
extragalactic radio sources, and the results of
\cite{2011arXiv1111.4478G} may be scaled down to the GC, due to the
high density of molecular clouds. A sketch of our model is provided in
Figure~\ref{jet}.
}

The ejection efficiency is the central idea of our model.  With $2 \% $ star formation efficiency (per dynamical time),  and a cloud life-time of 10 dynamical times, regulated by AGN disruption, one could form a  total of $10^4 $  stars  each of mass $\sim 20 \rm M_\odot$  (assuming a   top-heavy IMF).  For a more conventional IMF, one would form far fewer OB stars, perhaps only $\sim 100$, requiring
a correspondingly higher ejection efficiency. We infer that 
$\sim 1\%$  of OB stars need to be ejected in the event.  In this case, the total stellar mass ejected is $\sim 2000 \rm M _\odot$ and the associated gas mass ejected is  $4 { \times} 10^4 \rm M_\odot$,  or a few percent   of the GMC.

These numbers  seem to work  for the Milky Way Galaxy.  Let us assume that in one of these events, 
100 stars of average mass $10 \rm M_\odot$ are ejected at 300 km/s. Their total momentum flux is $ 6{\times}10^{43}$ g cm/s. Over $10^6$ yr, the momentum flux  ejected is $2{ \times}10^{30} \rm dynes.$
The SMBH at the GC  has an Eddington luminosity of $5{ \times}10^{44} $ergs/s (for mass $4{\times}10^6 M_\odot$) and we have assumed that it radiates at 10\% of the Eddington rate. If  the outflow were momentum-driven,
the ratio of AGN radiative pressure $L_{Edd}(4\pi r^2 c)^{-1}$ to mechanical wind pressure $\dot Mv(4\pi r^2)^{-1}$
is $L_{Edd}(\dot M v c)^{-1}=\eta c/v \sim 10,$
where $v$ is the wind velocity and $\eta\sim 0.1$ is the radiative efficiency of accretion.
The associated momentum flux (for optical depth unity, f=1) is 
$f L_{Edd}/c=1.7\times10^{34} f \rm dynes $ (actually observations require a factor $f \sim10$ to account for the $M_{BH}- \sigma$ relation \citep{2010ApJ...725..556S}).
In this case,  the ejection efficiency is $1.2\times10^{-4}/f. $ This seems reasonable even for $f_E\sim 0.1.$ 
%

The momentum transfer
is  determined by the ram pressure of the jet-driven expanding blast wave.   The increased surface area of the expanding  bow shock results in a substantial  boost in momentum-driving 
 \citep {Krause and Gaibler 2010}, more quantitatively demonstrated in \cite{2012arXiv1205.0542W}.
Additional ways of enhancing the momentum transfer  include
an energy-driven blast wave  (\cite{2003ApJ...596L..27K,   2012arXiv1204.2547F})
and  a non-isothermal dark halo
\citep {2012MNRAS.tmp.2997M}. Similar enhancements are required 
to explain the observed outflows, as recently advocated in \cite{2011ApJ...733L..16S}. 

We estimate the ejection velocity of a gas clump by assuming that a jet with power $10^{44}L_{44} \rm ergs/s$ drives a blast wave into a cloud
with initial
density $10^5n_5$~cm$^{-3}$ and radius  $5R_5\rm~pc$ 
(we ignore density stratification here). 
We emphasize that 
it is not actually the jet itself that accelerates clumps and induces star formation, but rather the associated  cocoon/shock wave.  Note also  that the tidal radius is roughly 
 $r_t ~\sim r_c (M_c / M_{BH})^{1/3} $ (e.g. (\cite {2011arXiv1112.4822M}).
  If the tidal radius is larger then the extent of the cloud{,} star formation is not affected by tidal forces,
  and self-gravitating cores are further stabilized.

The bubble radius evolves as 
$R_b=\left({ {5L}/({4\pi\rho_0})}\right)^{1/5}t^{3/5},$ and the cloud
is eroded over  a time 
  $R_b^{5/3}((4\pi/5)\rho/L)^{1/3}
  \sim 5{ \times}10^3 R_{5}^{5/3}(n_5/L_{44})^{1/3}\rm yr$, 
shorter than the free fall time. While 
star formation in the whole cloud abruptly terminates,
the diffuse gas of the cloud is compressed into a dense
  shell that cools on a very short timescale $t_c\approx 100 T_7^{1/2}
  n_5^{-1}$ years, where $10^7 T_7\rm K $ is the initial temperature of the
  shocked cloud gas. 
Any initial density inhomogeneities will then clump due to
thin shell and gravitational instabilities 
\citep{1983ApJ...274..152V}. 
At the edge of the cloud, the clumps should acquire a velocity of $v_b =
600 L_{44}^{1/3} n_5^{-1/3} R_5^{-2/3}$~km/s.
{The momentum of the shell when it reaches the edge of the cloud would be $10^{47}$ g cm/s, sufficient to explain the HVS observations, even if only a fraction of the gas forms stars.}The time scale difference between acceleration and fragmentation  is important:  the cloud will in fact be disrupted, but stars are formed in the process .

These clumps should  cruise  at basically
constant speed until they form stars{, because due to clumping and
density stratification the clumps  are much denser than their surroundings.}
If we assume a ten to hundredfold increase of the density in the shell,
compared to the average cloud density, we infer a free fall time of
several $10^4$ up to $10^5$~yrs. 
Assuming an exterior density of 100~cm$^{-3}$, as observed in
the hot gas today \citep{2003ApJ...591..891B}, the expanding shell would have to
accelerate to several 1000~km/s. This triggers Rayleigh-Taylor
instabilities, so that the clumps  disconnect from the shell, which
reforms at higher speed and continues to sweep away the
interstellar medium. 
{ Even at the present epoch, the observed jet might
  lead to some HVS ejections, as direct jet-cloud interactions should
  lead to molecular cloud acceleration up to $\approx 270$~km/s 
\citep{2012arXiv1208.1193Y}. 

HVS formation might also manifest an anisotropic 
geometry.}
There is a suggestion that the hypervelocity stars are found in two thin disk planes \citep{lu10}.
A disk-like geometry is indeed suggested by the numerical simulations of triggered star formation \citep{2011arXiv1111.4478G}.
Skewed jets are commonly observed \citep{2000ApJ...537..152K,2011MNRAS.414.2148L}, and it is hardly plausible that successive misaligned feeding disks would prefer the same plane of symmetry  
\citep{2006MNRAS.368.1196L}.



Another  manifestation of a recent AGN triggering episode may be the recently discovered Fermi gamma ray bubbles. The Fermi bubbles require $E_{bubble}= 10^{55} $ erg  or more in energy input into the  relativistic plasma.  The associated momentum is $ E_{bubble}/c= 3 {\times} 10^{44}$ g cm/s.  Over $10^6$ yr this amounts to a  rate of  $10^{31}$ dynes.  It is also roughly equal to the momentum flux in the ejected stars  for the usual SFE of 
$10\%$  of that in the total gas plus stars ejected). This seems reasonable if most of the energy injected is thermal.  In fact the required efficiency for driving the star-forming clumps is even smaller as the AGN wind model for the Fermi bubbles   \citep{2012arXiv1203.3060Z}
requires $\sim 10^{57}$ ergs injection in  a few $ 10^5$  yr and
 a jet power  of $\sim 10^{44}$ ergs/s.

\section{Cosmological implications}
One consequence of jet-induced clump acceleration leading to young star ejection is  that jet-induced cloud/star motions could retain some orbital memory of a central as opposed to a disk
injection mode. Another global consequence is that nuclear ejection would have happened more often in the past for our MWG, and much more vigorously in the past for galaxies with more massive BH than that of our MWG. 

Some 100 $ 10 
 \rm M_\odot$ stars ejected per $10^7$ yrs imply $10^5$ star ejections over the age of the MWG.  Let us allow a factor  of  $\sim 10$  for past enhanced activity, that is   some $10^6 $ stars or $10^7 \rm M_\odot$ are ejected from the MWG. These stars would become SNII in the intergalactic medium (IGM). The inferred mean metallicity generated  is $10^{-4}$ solar (since a L$_\ast$  galaxy of $10^{11} M_\odot$  accounts for solar yields).  This provides the observed IGM abundance floor. Moreover the  IGM abundance floor (which is indeed about this value from Ly alpha forest data) should be $\alpha-$enhanced.
The stars  would have travelled a distance (300- 3000) km/s { $\times$} 6 Myrs or  1.8-18 kpc.  Over $3{\vg \times}10^{7}$ yrs,  this amounts to 9-90 kpc. This results in IGM enrichment on halo and on group scales. This is about what might be expected, for an enrichment of the surrounding MWG  IGM at z=0, and will provide ubiquitous enrichment  for the IGM at $z\sim 2$  when AGN activity peaks and distances are smaller.

Of course,  the number of events depends on the gas refuelling rate. Since the gas accretion and SMBH fuelling rates were  higher in the early MWG, one ejects even more stars. A rough estimate comes from scaling the SFR, which is  fed by cold gas accretion and  also scales with the BH accretion rate by a ratio of about 1000 (see \cite{2012arXiv1204.2824M}).  The early SFR is enhanced by about a factor of 10 in order to account for disk chemical evolution, eg. \cite{2009A&A...505..605C}.
This gives a factor of approximately  10 boost in OB star ejection relative to a constant rate, with corresponding implications for the time-dependence of the chemical evolution of the IGM.  The fact that the triggering  mode is more efficient at high $z$ suggests  that any surviving high velocity stars  from early episodes should be biased to lower metallicity.

%
%

A  time delay of 50-100 Myr has been measured for several  hypervelocity stars apparently ejected from the GC \citep {2012ApJ...754L...2B}{\vg .}
 These represent  only a small fraction of the hypervelocity stars. Our model predicts that  only $\sim 1\%$ of the newly formed 
jet-induced stars are ejected promptly.
It is possible that  a significant fraction  of the newly formed massive stars may have  massive binary companions, leading to  
a  time delay of   $4{ \times}10^7$ yr  or longer,  if the 
binary companion  exploded as a core collapse supernova or as a 
 prompt SNIa single degenerate core  \citep{arXiv:1206.0465}. 

In summary, we have  argued that  many of the observed hypervelocity stars were generated by the interactions of an AGN jet-driven cocoon  from the central black hole with a dense molecular cloud.  There are
broader implications for nuclear star formation, the enrichment of the intergalactic medium  and the possible origin of the Fermi bubbles.

}

\begin{acknowledgements}
JS and YD acknowledge support by the ERC project "Dark Matters", MK by the cluster of excellence ``Origin and
Structure of the Universe'' (www.universe-cluster.de), and VG by the Sonderforschungsbereich SFB 881 ("The Milky Way System", subproject B4) of the German Research Foundation (DFG).
We  thank  Zachary Dugan and Rosemary Wyse for discussions.
\end{acknowledgements}

\end{document}